\documentclass[
aps,%
12pt,%
final,%
notitlepage,%
oneside,%
onecolumn,%
preprintnumbers,%
nobibnotes,%
nofootinbib,%
superscriptaddress,%
noshowpacs,%
centertags]%
{revtex4}
\usepackage{graphicx}

\def\la{\mathrel{\mathpalette\fun <}}

\def\fun#1#2{\lower3.6pt\vbox{\baselineskip0pt\lineskip.9pt
\ialign{$\mathsurround=0pt#1\hfil##\hfil$\crcr#2\crcr\sim\crcr}}}
\def\msun{M_\odot}

\let\jnlstyle=\rm
\def\refjnl#1{{\jnlstyle#1\,}}
\def\aj{\refjnl{Astron.J.}}
\def\apj{\refjnl{Astrophys.J.}}
\def\azh{\refjnl{Astron.Zh.}}

\def\mnras{\refjnl{Mon.Not.Roy.Astron.Soc.}}

\def\apjl{\refjnl{Astrophys.J. Lett.}}
\def\apjs{\refjnl{Astrophys.J. Suppl.}}

\def\aap{\refjnl{Astron.Astrophys.}}

\def\prd{\refjnl{Phys.~Rev.~D}}

\def\pasp{\refjnl{The Publications of the Astronomical Society of the Pacific}}

\def\BE{\begin{equation}}
\def\EE{\end{equation}}
\def\BA{\begin{array}}
\def\EA{\end{array}}
\def\BAN{\begin{eqnarray}}
\def\EAN{\end{eqnarray}}

\begin{document}
\preprint{\vtop{
{\hbox{IPMU09-0044}\vskip-0pt
}
}
}

\selectlanguage{english}

\title[Mirror World]{Notes on Hidden Mirror World}


\author{\firstname{S.~I.}~\surname{Blinnikov}}
\email{sergei.blinnikov@itep.ru}
\affiliation{%
Institute for Theoretical and  Experimental Physics (ITEP),
117218, Moscow, Russia
}
\affiliation{
  IPMU, University of Tokyo,
5-1-5 Kashiwanoha, Kashiwa, 277-8568, Japan
}%

\begin{abstract}
A few remarks on Dark Matter (DM) models are presented.
An example is Mirror Matter which is the oldest but still viable DM candidate, perhaps not in the
purest form.
It can serve as a test-bench for other analogous  DM models, since
the properties of macroscopic objects are quite firmly fixed for Mirror Matter.
A pedagogical derivation of virial theorem is given and it is pointed out that
concepts of virial velocity or virial temperature are misleading for some cases.
It is shown that the limits on self-interaction cross-sections derived from observations
of colliding clusters of galaxies are not real limits for individual particles if they
form macroscopic bodies.
The effect of the heating of interstellar medium by Mirror Matter compact stars
is very weak but may be observable.
The effect of neutron star heating by accretion of M-baryons may be negligible.
Problems of {\sc macho}s as Mirror Matter  stars are touched upon.
\end{abstract}

\maketitle

\section{Introduction}
\label{intro}

Dark Matter (DM) remains one of the main puzzles in modern
physics and astrophysics.
In this paper I have collected a few notes on
one DM candidate, namely, Mirror Matter.
I do not attempt to give a full review on the subject: only
some comments on Mirror Matter properties from the point
of view of astrophysics are put forward.
The comments may be useful for particle physicists working
on new models for DM, e.g. from hidden sector, because Mirror Matter
serves as an example of the model, where not only particle properties,
but also the properties of macroscopic objects are quite firmly fixed
and well known.

\section{Dark Matter manifestations}
\label{DMmfst}

If Newton-Einstein's law of gravity is not modified at large scales,
then DM manifests itself in many circumstances.

First, in 1933, Fritz Zwicky has discovered  virial paradox
in Coma cluster of galaxies \cite{Zwicky1933}:
velocities of galaxies, $u_{\rm gal}$, are on average
much higher there than can be expected from an estimate of gravitational
potential using only the mass of visible matter, $M_{\rm vis}$:
$$
 u_{\rm gal}^2 \gg u_{\rm vir}^2  \sim \frac{G_N M_{\rm vis}}{R} \; .
$$
We will discuss below more accurate formulation of the virial theorem, and
dangers of using notions like the  ``virial velocity'' ($u_{\rm vir}$
in the above expression) in some situations.
But here the discrepancy is real.
Zwicky immediately suggested the presence
of ``dunkle Materie'' \cite{Zwicky1933}, i.e. the Dark Matter to remove
the discrepancy.

Very soon, just after the paper by Einstein \cite{Einstein1936} on gravitational lensing (already discussed by Chwolson
\cite{Chwolson1924} for ordinary stars),
Zwicky has proposed that the presence of DM can be
tested by the effect of gravitational lensing
\cite{Zwicky1937pr1,Zwicky1937pr2}:  this what we call
now macrolensing.
It is interesting to note that the idea was suggested to Zwicky by
Vladimir Zworykin -- the Father of television.
(Zwicky writes in \cite{Zwicky1937pr1} that Zworykin has got the idea
from R.W.Mandl, but
I could not find a reference to any paper by  R.W.Mandl on that.)
Many wonderful examples of macrolensing are known now, like multiple
images of distant
quasars and galaxies, and they all show that the mass of the lens
(an intervening cluster of galaxies) must be much larger than
the mass of visible component in the lens (again, if general
relativity gives the true
description of gravity at relevant distances).
The lensing on early stages of research was also discussed by
Tikhov \cite{Tikhov1937,Tikhov1938}.
Other tools, important especially for compact dark objects,
like microlensing were developed later
\cite{Refsdal1964,Liebes1964,Byalko1969,Byalko1970,GurZybSir1996,GurZybSir1997r,GurZybSir1997e},
see  a review in \cite{Zakharov2008}.

Very important for DM studies is the dynamics (e.g. rotation curves)
of galaxies,
also first proposed by Zwicky in 1937 \cite{Zwicky1937apj}.
Real measurements of  missing mass growing with radius in M31 (Andromeda)
galaxy have been performed already at that time \cite{Babcock1939}.
But Zwicky's seminal ideas and papers were almost forgotten, and
rediscovery of DM occurred only four decades later, in 1970s.
Then it was much more popular in USSR, than in the West, where it was
accepted only after strong resistance, see the review by Einasto
\cite{Einasto2009}.

New evidence in favor of DM came from discovery of
huge amount of X-ray emitting  gas in clusters of galaxies.
Mass of this gas is order of magnitude larger than the total visible mass
of galaxies, but still not sufficient to  resolve the virial paradox.
The temperature of the gas, $T  \sim \mbox{a few}$~keV, corresponds to
the velocity of protons of the same order as relative velocities of
galaxies $u_{\rm gal}$ in a cluster.
The hot gas cannot be trapped by the gravity of visible baryons, and this
also points to the need of DM.

More detailed history of DM discovery see in
\cite{vdBergh1999,Biviano2000,Roberts2008,Einasto2009}.

Probably, the most  important manifestation of DM is its
influence on structure formation.
A homogeneous universe in Friedmann solutions
contains no galaxies or clusters -- no structures.
Matter in Universe was initially almost uniform, with small perturbations
left after inflation
 \cite{Sta1979,Sta1980,MukhChib1981,Sato1981,Sta1982,Guth1982,Sta1983,Sta1985}
(or another initial stage, e.g.,  in cyclic scenarios \cite{CraHertTurok2007}).
After the recombination
gravity enhances the density contrast,
leading to the formation of structures when the self-gravity
of an over-dense region becomes large enough to decouple the blob of matter from
the overall expansion.
Cosmic Microwave Background (CMB) fluctuations tell us that
perturbations in density of baryons at epoch of recombination at redshift
$z = z_{\rm rec} \sim 10^3$ were on the level of $10^{-5}$.
They can grow only as the scale-factor $a \propto (1+z)$ according
to the Lifshits theory of growth of small perturbations in an expanding universe.
Thus, now, at $z=0$, we would get not more than 1\% fluctuations of density,
contrary to the observed structure.

DM particles can have larger amplitude of density perturbations and they begin
to form structures earlier, while the baryonic
matter is overwhelmed by photon pressure until the time of recombination.
If DM dominates in gravity,
the ``dark matter boost'' is provided for structure formation in visible matter,
see e.g. \cite{GurZyb1995r,GurZyb1995e}.

\section{Dark Matter Candidates and Mirror Matter}
\label{sec:DMCand}

Current estimates of the fraction of DM  in the average density of Universe
are around 25\% (assuming
standard gravity), while baryons cannot contribute more than a few percent as found
from abundances of light elements in primordial nucleosynthesis \cite{Steigman2007} and
analysis of CMB \cite{WMAP2007}.

If Dark Matter is real and non-baryonic, the most acute question is: ``What is DM made of''?
A clear and consize  introduction to possible DM candidates is foind in G.Raffelt's reviews \cite{Raffelt1998,Raffelt2004}.
From the last one, at ``The Dark Universe'' conference, Munich, May 2004, one can point out a few
probable models for DM.

 {
    \begin{itemize}
    \item[--] Supersymmetric particles ---  neutralino for heavy WIMPs
     (weakly interacting massive particles), or   gravitino for lighter ones
          and their clusters.
    \item[--] Axion-like particles and objects made of those particles.
    \item[--] Mirror Matter particles and objects \cite{KOP1966}.
This is not so popular now as a DM model,
          but actually it is the oldest proposed candidate.
    \end{itemize}
}

In spite of years of intensive search for WIMPs,
nothing yet is discovered in laboratory!
That is why Mirror Matter  $\equiv$ MM  is still a viable Dark Matter candidate:
MM particles in the simplest MM model interact with ordinary matter only gravitationally
(if we do not involve weak electromagnetic coupling like in \cite{Foot2008}),
and they can not be detected in laboratory experiments.

Neutralino and axions have electromagnetic interactions (on weak scale).
MM in pure form has no electromagnetic
 (or other standard model) interactions
with Ordinary Matter $\equiv$ OM, only  gravity.
MM has analogs of all ``our'' interactions  within Mirror sector.
There are also models with slightly ``broken'' Mirror symmetry
which have weak  electromagnetic coupling to OM.

Hypothesis on existence of invisible (or Dark) Mirror Matter
was put forward by Kobzarev, Okun and Pomeranchuk in 1966
in the  paper ``On the possibility of observing mirror particles''
\cite{KOP1966} after the discovery
of P- (1956)  and CP-parity non-conservation (1964), when the mirror asymmetry of visible
matter had become obvious.
They have considered possible consequences of existence of  matter  which
is mirror relative to our matter.


The history is often distorted in modern revues: Lee and Yang \cite{LeeYang1956}
 introduced a concept of right-handed particles,
 but their  ``R-matter'' was not hidden! -- as some of recent papers claim, e.g. \cite{Feng2008}.
An exact citation from  \cite{LeeYang1956} reads:

``\ldots the interaction between them is not necessarily weak. For
example, $p_R$ and $p_L$ could interact with the same
electromagnetic field and perhaps the same pion field.
They could then be separately pair-produced, giving
rise to interesting observational possibilities''.

Kobzarev, Okun and Pomeranchuk \cite{KOP1966} have shown that common
electromagnetic and strong interaction of O and M particles are not allowed,
 this would contradict experiments.
They demonstrated that only gravity and very weak interaction were possible between
the two kinds of particles.
So, it was shown that Mirror Matter lies in the hidden
sector for the first time in \cite{KOP1966}.
The hidden world of Mirror Matter  has the same microphysics as the ``visible'' one.
The term ``Mirror Matter'' was  introduced first time also in \cite{KOP1966}:
it was taken by the authors from the book by Lewis Carroll
``Through the Looking-Glass''.

Another  important paper by L.B.~Okun,
 ``On the search for new long-range forces'',
\cite{Okun1980r,Okun1980e}
discussed other DM forms.
The idea of MM search by the effect of gravitational lensing was also put forward
in \cite{Okun1980r,Okun1980e}.

Our papers with M.Yu.Khlopov
``On possible effects of `mirror' particles''
\cite{BliKhlo1980,BliKhlo1982r,BliKhlo1982e,BliKhlo1983r,BliKhlo1983e}
began systematic studies of astrophysics of  MM   as a form of  DM.

After \cite{KOP1966} about 300 papers have been published
(among them e.g. Z.Berezhiani,
R.Foot  were very active, e.g. \cite{Berezhiani2001,Berezhiani2005,Foot2004}), an extensive review of
the literature on the subject  is written by Okun \cite{Okun2007}.
More recent (brief) review by Z.K. Silagadze
has an intriguing title ``Mirror dark matter discovered?'' \cite{Silagadze2008}.
However, to remove the question mark in this statement would be premature:
Mirror world is still hidden from us.

\section{Virial Paradoxes}
\label{VT}

The discovery of DM by Zwicky  was based on virial theorem. Here I derive the virial
theorem and show that for some DM candidates its application is not trivial.
We have for energy $E$ of  a stationary state for a system with Hamiltonian $H$
$$
   \delta E_{\mbox{tot}} \equiv \delta {\cal E} = \delta \langle H \rangle=0
$$
in the first order of perturbation of its wave function $\delta\psi$, i.e. the variational principle
in quantum case.
This is a starting point for seeking for the estimates of the ground state energy
of quantum systems using trial approximation functions for $\psi$ in standard applications
of the variational principle.
More interesting for us is the virial theorem.
{Compare the text below with derivation of virial theorem by  Fock 
\cite{Fock1930} in a quantum system,   
here I present it for a particular  N-body problem with Coulomb potential.}

Let us take a perturbation of the form:
$$
  \psi+\delta\psi = \alpha^{3N/2} \psi(\{\alpha \vec{r_i}\}),\quad
  i=1,\ldots,N,
$$
so the wave function is uniformly changed for all $3N$ space coordinates.
The coefficient $\alpha^{3N/2}$ is from normalization to unity.
Note, that $\alpha > 1$ here corresponds to the {\em compression} of the whole
system.

Schr\"{o}dinger equation:
$$
    \imath \hbar \frac{\partial\psi}{\partial t} = H \psi
$$
with
$$
  H= -\sum_j\frac{\hbar^2}{2m_j}(\nabla^2\psi) + U\psi
$$
gives after averaging
$$
  \langle H \rangle= \langle E_{\rm kin} \rangle + \langle U \rangle.
$$
Now we have after the perturbation:
$$
  \langle{E_{\rm kin}}\rangle \rightarrow \alpha^2 \langle{E_{\rm kin}}\rangle
$$
because for non-relativistic (NR) particles, $E_{\rm kin} \propto p^2 \propto 1/\lambda^2$, i.e.
$$
 \frac{\partial^2\psi(\ldots \alpha x_i \ldots) }{ \partial x_i^2} = \alpha^2
 \frac{\partial^2\psi}{\partial x_i^2}.
$$

The variation of $U$ depends on the law of interaction.
For Coulomb (and Newton!) interactions
$$
 \langle U \rangle \propto \int\sum_{i\neq k} \psi^* \frac{1}{r_{ik}} \psi \,
 d^N\vec{r} \rightarrow \alpha \langle U \rangle.
$$

Thus
$$
  \langle H \rangle \rightarrow \alpha^2\langle E_{\rm kin} \rangle + \alpha\langle U
  \rangle,
$$
and variation of this gives
$$
  \delta\langle H \rangle = 2\alpha\delta\alpha\langle E_{\rm kin} \rangle +
  \delta\alpha\langle U  \rangle =0,
$$
so with $\alpha=1$ for unperturbed state we find
$$
  2\langle E_{\rm kin} \rangle + \langle U \rangle = 0.
$$
This is the virial theorem for atomic Coulomb potential and for globular
stellar clusters, and for clusters of galaxies as well!
See the use of Schr\"{o}dinger equation for stellar dynamics
in \cite{WidKai1993,Johnston2009}. 

For all those systems (NR atoms or plasma, NR stars and clusters)
$$
  {\cal E} = \langle E_{\rm kin} \rangle + \langle U \rangle = - \langle E_{\rm kin} \rangle ,
$$
so the loss of total energy ${\cal E}$ corresponds to the {\em growth} of the
kinetic energy $\langle E_{\rm kin} \rangle$.
The same is true for the internal energy of matter if it is in the form of
kinetic energy of particles.
We should remember that we derived this based on the stationary quantum state
which corresponds to an orbit in a classical cluster.
The cluster must be in a relaxed (so-called ``virialized'') classical state,
i.e. it should be considered on a time-scale much longer than the orbital period.
Otherwise, e.g. on a stage of violent relaxation, the virial theorem cannot be applied.


Now let us consider {\em non-interacting} particles in a potential well.
From Schr\"{o}dinger equation for NR particles (and also for classical objects!),
if $U \propto r^k$,
we find in the same way as above:
$$
  2\langle{E_{\rm kin}}\rangle - k \langle{U}\rangle=0.
$$
If $k=-1$ we have Coulomb and Newton laws, for  $k=2$ --- a harmonic oscillator.

The case $k \to +\infty$ corresponds to a hard reflecting
wall, then $\langle U\rangle$ tends to zero relative to
$\langle E_{\rm kin}\rangle$, but the variation of  $\langle U\rangle$
is always of the same order as the variation of
$\langle E_{\rm kin}\rangle$.
Since the `force' on the particle is $-\nabla U$, we define pressure $P$
(i.e. the `force' on unit area, $A=1$, of the wall) so as
$$
  \delta\langle U  \rangle = P A\delta x = P \delta V.
$$
For our variations $V \rightarrow V/\alpha^3$,
and when we express
variations of $H$ again through $\alpha$, we get
$$
  2\langle{E_{\rm kin}}\rangle - 3PV=0 ,
$$
i.e. the same relation as in classical derivation when the moments of equation of motion are
averaged in time  (see e.g. Kubo's textbook on statistical mechanics).

Thus, we have for pressure in NR case:
$$
  P = \frac{2}{3} \varepsilon
$$
so
$$E_{\rm kin} = E_{\rm thermal}= \frac{3}{2}\int P dV \; .
$$

Important for us is also an extremely relativistic (ER) case \cite{Fock1930}:
$E_{\rm kin} \propto p \propto 1/\lambda$, then

$$
  \langle E_{\rm kin} \rangle + \langle U \rangle = 0.
$$

For ER case:
$$
  P = \frac{1}{3} \varepsilon
$$
and
$$
E_{\rm kin} = E_{\rm thermal}= {3} \int P dV\; .
$$
and from virial theorem in both NR and ER cases we have got
$$
  3 \int P dV + \langle U \rangle = 0 \; .
$$
Of course this relation is valid for arbitrary
$P=P(\varepsilon)$ and its derivation can be found in standard textbooks.

Now let us discuss why it is dangerous sometimes to use the
virial theorem  and  to rely on $T_{\rm vir}$ and  $v_{\rm vir}$ for DM particles with
long-range interactions.

For $U \sim - G_{\rm N}M^2/ R $,
omitting all coefficients of order unity, we find that
pressure and density in the center of a spherical
cloud of mass $M$ and radius $R$ are:
$$
P_c \simeq \frac{G_{\rm N}M^2}{R^4}, \;\rho_c \simeq \frac{M}{R^3} .
$$
and we have
$$
P_c \simeq G_{\rm N}M^\frac{2}{3}\rho_c^{4/3}.
$$

So, if we have a classical ideal plasma with $P={\cal R}\rho T/\mu$,
where ${\cal R}$
is the universal gas constant,
and $\mu$ --- mean molecular mass, we get
$$
 T_c \simeq \frac{G_{\rm N}M^{2/3}\rho_c^{1/3} \mu }{ {\cal R}}.
$$

With $\mu \simeq 1$ for hydrogen-helium fully ionized plasma we get for the Sun $T_c \simeq
10^7\; \mbox{K} \simeq 1$ keV.
This is OK for ``virial'' velocity $ v^2_{\rm vir} \sim \varphi \sim r_g/R_\odot \sim 10^{-6} $
{(although $\sqrt{2}$ less than for neutrals!)}.
The temperature $T_c$ may be called a ``virial'' $T_{\rm vir}$ for ions, but not for electrons,
the latter have much higher velocity than ``virial'', since they have the same temperature but
much smaller mass.

The discrepancy is much larger in degenerate stars.
E.g. in Sirius B, with its central density higher than $10^7 \mbox{g cm}^{-3}$,
we have velocity of electrons $\sim c$
(due to high Fermi momenta in degenerate electron gas)
and velocity of ions tends to zero.
They are hot enough to form a classical ideal plasma,
but $T$ of course is quite different from the classical virial value.

This has direct relevance to DM studies.
One can imagine a DM candidate like a neutrino of a few eV mass.
It will determine the mass of a cluster or super-cluster of galaxies \cite{BKNov1980}, but
 $T_{\rm vir}$  loses its sense when degeneracy becomes important.
Situation  with long-range forces
(like, e.g.,  in models for hidden charged Dark Matter  \cite{Feng2008}) is analogous
with ordinary plasma, and neither virial velocity,
nor virial temperature are appropriate concepts for
DM candidates of this sort.

This paradoxical conclusion does not mean that the virial theorem is not valid.
Simply its standard derivation,  and the formula $3 \int P dV + \langle U_g \rangle = 0$, where
$U_g$ is the gravitational potential energy,
mask the fact that all other types of potentials must be accounted for.
Inside a star electrons always have velocity much higher than $u_{\rm vir}$, either due
to temperature which must be equal to that of ions, or due to high Fermi momentum in
a cold star. So electrons tend to fly away from the gravitational potential well,
and immediately a very small polarization of plasma occurs.
Electrons feel electrostatic field
which accelerates them much stronger than gravity because of their small mass.
Ions feel electric force of the opposite sign which equilibrates the gravity.
With account of the
electrostatic potential, added to the gravitational potential in the virial theorem,
the particles of different sign have velocities of the right order.

The concept of pressure $P$ has to be used with care as well.
In a degenerate dwarf  the pressure of electrons is much higher than the pressure of ions,
but  it cannot be said in general that  $P$ describes the transfer of momentum in collisions of electrons
and ions, like in classical fluids: there are no individual collisions for degenerate
particles.
Electrons cannot lose their momentum: the lower states are all occupied.
They are in steady states  (without collisions) in a potential well of electric field
with a size of star.
Ions may form an ideal gas, or liquid, or even a crystal and they are supported against the gravity
by the {\it volume} force of the same electric field.
So $P$ enters here  not like a   {\it surface} force on a ``brick wall'',
but just as a flux of momentum per unit time
mediated by electric field.

\section{Accretion onto M-objects}
\label{Accretion}

Some popular DM candidates like neutralino are not quite ``dark''.
They may, e.g., annihilate and produce gamma-rays. Actually, this is
one of the most promising way to identify them in space.

Pure MM manifests itself via gravity only.
The only source of radiation of O-photons by an M-star
is the accretion of the O-matter. The term ``accretion''
denotes here not only the capture of matter by a star,
but the whole complex of the processes of gravitational
interaction of the M-star with the surrounding O-medium.
MM stars may accrete ordinary matter and
one may hope to detect a signal of an unusual accreting object.
 Early proposals to search MM objects along these lines have been put forward in
\cite{BliKhlo1980,BliKhlo1983r,BliKhlo1983e}.

Standard accretion of gas on gravitating center is described by so called
Bondi--Hoyle--Lyttleton (BHL) theory,
 see  the review by R.G.~Edgar \cite{Edgar2004} for references.
Details of accretion of non-interacting particles on a gravitating center
are considered in \cite{Ill1985r,Ill1985e}.

\begin{figure}
\setcaptionmargin{25mm}
\onelinecaptionsfalse
\includegraphics[width=0.5\linewidth]{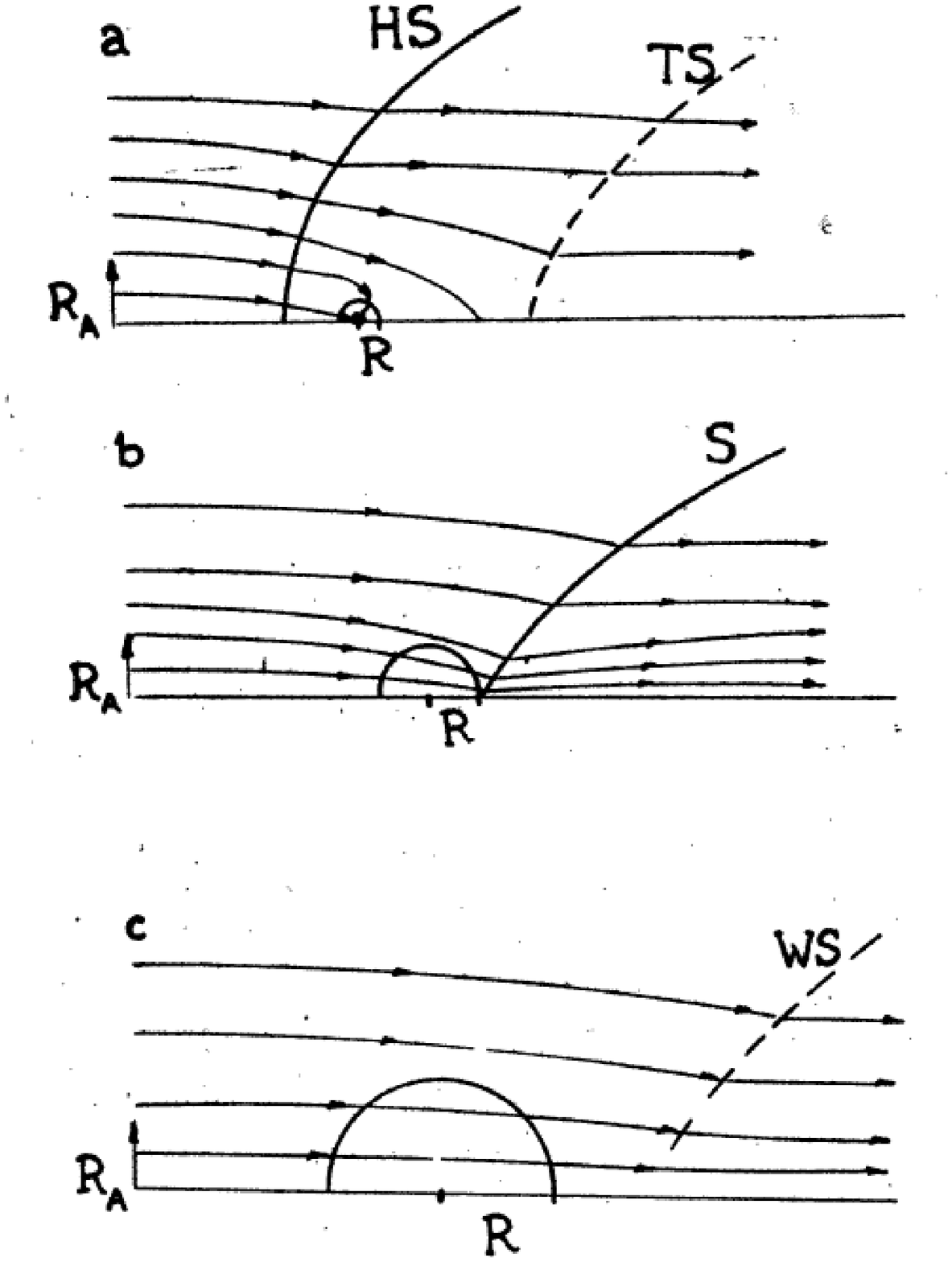}
\captionstyle{normal}
  \caption{{  Possible regimes of accretion onto an M-object moving
supersonically through O-matter from \cite{BliKhlo1980}:}
\newline
a) $R \ll  R_A \, ;$ HS denotes a head (bow) shock, TS --- a possible
trailing shock; {  the matter is partially captured and cools down in the potential
well of M-star}.
\newline
b) $R \approx R_A$; S is a shock front; {  the matter is virtually not
captured}.
\newline
c) $R > (u/u_s)R_A$; WS denotes a weak shock (in general,
presumably, may not form at all); {  the matter flow is only slightly
perturbed, no capture}.
}
 \label{MMaccr}
\end{figure}

According to the classical theory of accretion,
the gravitating body of mass M effectively captures the
matter within the accretion radius
$$
R_A = 2G M /u^2 \; ,
$$
where $u$ is the velocity of the body's motion through the
medium in the supersonic case, or the sound speed in the
medium, $u_s$, in the subsonic case.
We may use very crude estimates here, so
the sound speed is roughly related to the temperature by $k T= m u_s^2$.
The square of $R_A$ gives the cross-section $ \propto \pi R_A^2 $ and the result is
essentially the same as for Rutherford scattering (replacing $GM$ by $Ze$).

For $u> 10$ km/s the relative motion is supersonic,
since $u_s \simeq 10$ km/s for the temperature of the medium $T \sim 10^4$~K.
Supersonic accretion onto M-star was considered qualitatively by us \cite{BliKhlo1980,BliKhlo1983r,BliKhlo1983e}.
If the radius of a star $R$ is much smaller than $R_A$, then accretion
must be similar to accretion onto a black hole or a neutron star but there is no trapping
surface.
For larger radii  there are  different regimes possible,
see Figure~\ref{MMaccr}, depending on ratio
of the star radius $R$ and the accretion radius $R_A$.

For a very large  radius $R$ there is hardly any hope to detect the emission
of the perturbed gas.
So the case $ R \ll R_A $ which obtains easily for normal stars, and is especially good for
white dwarfs and neutron stars, is worth to study.
This case must be close to the BHL accretion.

For such a case
G.~S.~{Bisnovatyi-Kogan} at al. 
\cite{BKKKLSh1979} have found a self-similar
solution for a gas with constant adiabatic exponent $\gamma$. This solution has
some singularities and numerical results \cite{Edgar2004} show a different picture,
more similar to our Figure~\ref{MMaccr}a.

Using formulae for accretion rate one can estimate the amount of  mixing of O- and M-matter.
From
$$
R_A = \frac{2G M}{u^2 } \approx 3\times 10^{14} \frac{M}{\msun} \frac{1}{u_6^2} \; \mbox{cm} ,
$$
we get $R_A$ about tens a.u. for $M = \msun$.
If $\rho \sim m_p n  $ is the density of the surrounding medium
then maximizing the classical expression  for a rate of accretion
we get
$$
\dot M = \pi R_A^2 \rho u  \sim 10^{12}\left(\frac{M}{M_\odot}\right)^2  \frac{n}{u_6^3}  \;
\mbox{g/s}
$$
and for $n \sim 1 \; \mbox{cm}^{-3}$ a star accretes  $3 \times 10^{28}/{u_6^3}$ g in
$10^{10}$ yrs, a bit heavier than Earth mass.

Luminosity of infalling gas is approximately given by
$$
L \sim \dot M \varphi_0 \; ,
$$
where $ \varphi_0$  is the characteristic value of the gravitational
potential in the place of the stopping of the infalling matter.
If the radius of an accreting M-star $R$ is of order of the neutron star radius,
then $R \ll R_ A$, $\varphi_0 \sim 0.1 c^2$ and $L$
will approach a few percent of Solar luminosity for a single M-star.
Luminosity may be much higher in a binary of O-M stars, however, the
probability to find such a binary is very low \cite{BliKhlo1983r,BliKhlo1983e}.

It is clear that to discover a MM star by its accretion of ordinary matter
is extremely hard.
It is hard even for a single old neutron star made of ordinary visible matter.
Nevertheless, the chances are not absolute zero.
Crude estimates of radiation fluxes produced by accretion
onto a MM star (or any other hidden matter star) can be obtained
as follows.

Particle number density can reach $n \sim 10^6$~cm$^{-3}$ in molecular clouds.
It can be higher in a binary in a young stellar cluster.
Then for a MM neutron star one gets up to $L \sim 10^{38}$ ergs/s at
$n \sim 10^6$~cm$^{-3}$.
For a MM white dwarf the luminosity $L \sim 10^{36}$ ergs/s,
but the number of MM white dwarfs should be much larger.

It is hard to predict the spectral distribution without detailed
computations.
The hardest part of the spectrum should be determined by the stopping
radius (if there is no strong non-thermal tail).
For a neutron star it may go to X-rays.
It might be softer for a white dwarf.
But in both cases there is \textbf{no}
solid surface for O-matter in the beginning of accretion.
Later, after accumulation of some
amount of ordinary matter, such a surface may build up,
and spectrum will change.

The soft part of the spectrum may be determined by a total
area of excited region and by coherent radio emission.
For a black body and a surface of $4\pi R_A^2$ this implies
$T_{\rm bb} \sim $~a thousand K, and
a near IR wave-range for the maximum for this high $L\sim 10^{38}$  ergs/s.
But for $L \sim 4\times 10^{31} \sim 0.1 L_{\odot}$
the Wien peak is already in sub-mm range.

If we have  a sensitivity of 3 mJy then to detect an object located at 100 pc
we need it to have a luminosity not less than
$L_\nu\sim 10^{16}$  ergs/(s $\cdotp$ Hz). Thus, an instrument like RadioAstron
may detect an accreting
MM white dwarf if only  $\sim 10^{-3}$ of its weak
luminosity at $n \sim 1$~cm$^{-3}$ ($ \sim 0.001 L_{\odot}$) goes into radio
(for $\nu< 25$ GHz).
Or  much smaller fraction: $10^{-10}$ --- in case of $L\sim 10^{36}$ at $n \sim 10^6$~cm$^{-3}$.
This is quite plausible.

\section{Heating of neutron stars by MM}
\label{NS}

Now let us consider an opposite case:
accretion of MM onto an O-star.
It should  have quite similar patterns of flow as in
Figure~\ref{MMaccr} changing the flavours of particles accordingly.
Effects of trapping MM by O-stars were discussed in our old papers
\cite{BliKhlo1980,BliKhlo1983r,BliKhlo1983e}.

An interesting effect is discussed in  paper~\cite{Sandin2008}:
one can think that neutron stars capturing mirror matter could be heated 
at the order of 100 MeV 
per mirror baryon.
The cause of the heating can be ascribed to the necessity to emit
the excessive binding energy:
after accretion the final state has more mirror baryons than the initial state,
the number of ordinary baryons is constant, but the energy per ordinary baryon is
lower. Effectively the O-baryons  fall into a deeper potential well.

I show that the number of 100 MeV per mirror baryon in heating  is a strong overestimate.
During the process of accretion of M-baryons the matter in neutron
star really goes down into a deeper potential well.
But it is compressed due to slow accumulation of MM, and  this process is purely
adiabatic for O-matter and there is {\bf no} heating in O-sector. All
heating and emission and restoration of energy balances occur
in mirror-sector.  I propose several simple Gedankenexperiments illustrating that.

Let us have an O-neutron star and cover it with an ideally
reflecting surface. Let this star be ice cold. Let us call the baryons
in this star ``blue'' ones. Let  ``red'' (ordinary matter) baryons fall
slowly onto this star. Of course they will emit radiation, but let the
accretion rate be slow
enough for them to cool down to the same ice temperature. The
reflecting surface between blue and red baryons does not allow the
interiors to absorb any radiation. The total mass of ``red matter''  grows
together with the gravitational
binding energy, total energy of ``blue matter'' decreases, the
blue interiors do compress indeed, but they remain ice cold!

Now replace the red baryons with mirror matter and we need no
reflecting surface at all, because the radiation of mirror photons
(which keeps all energy balances) is harmless
for blue baryons. There is no effect of heating.

When one compresses  gas it heats up - but only if it is already hot,
$T>0, S>0$. If one compresses adiabatically then $T$ grows, but $S=\mbox{const}$.
If we lose heat, then $T$ may still grow, but $S$ goes down (like in
classical stars). Here I always mean $S$ per unit baryon.

But now we discuss the case of cold objects, $T=0, S=0$. Whichever strong
compression, $T$ does not grow if $S=\mbox{const}$.
It remains zero. To increase $T$ one must
have a strong non-adiabatic process, because $S$ must grow. Slow
accretion of mirror
baryons is an adiabatic process for O-neutron star, hence no growth of
$T$ whichever deep potential well develops.
One must produce a shock wave, or something like this, to heat it up.

Let us consider a charged ball of mass 0.1 kg falling in terrestrial
gravity and zero electric field $\mathbf{E}=0$ into a potential well  (a dielectric
hole) from 1 m. It must emit around one J until it comes to rest and
to a cold state at the bottom of the well.
Now let us switch on $\mathbf{E}\neq 0$ of such a strength that the
potential well is much deeper, say 100 J.
Now, after the falling, the ball must emit  100 J.
Finally, let us take
the  situation similar to an  accreting neutron star:  let us have the ball already
at the bottom of the well in case of $\mathbf{E}=0$  (this is our neutron star
and zero admixture of MM) and then
let us start increasing  $\mathbf{E}$ slowly (adiabatically!) -- this is the analog of  slow accretion of MM
particles and adiabatic change of gravity
-- the ball comes to a very deep potential well of 100 J.
But it emits virtually nothing:
it feels only adiabatic compression by electrostatic force,
and if its temperature was $T=0$ K
initially, it will remain so forever!
The same must be for the neutron star accreting MM.

If the reader does not like the electric field, one can do the same
Gedankenexperiment in a potential well with artificial gravity
(say in an accelerated laboratory, changing the force of engines
adiabatically, i.e. slowly).

To make my counterexamples extreme let us take one O-atom, say
of H and  put it inside a cloud of rarefied M-matter.
 Let the M-cloud eventually to collapse to a star,
and then to an M-neutron star with our atom in its center. It is still cold
but it is in a deep potential well,
like a  100 MeV/baryon  solely due to
the gravity of M-baryons. We need this order of  energy to pull it out of the well.
Do we need any process for our atom to emit 10 \% of its mass?
No, it could not do so. It was never excited. The
energy was surely emitted by M-baryons
in their M-photons, neutrinos etc. because they had to stop during accretion.

This is clear also in quantum language: any system in a ground state
will stay in this state when the potential changes adiabatically.
There are no excitations, no heating unless there are non-adiabatic
perturbations.

That is why the Hawking radiation of black holes is so weak (if any):
the adiabaticity of virtual modes can be violated only for wavelengths
of order $R_g$, hence  such a low $T$ of a BH (due to Wien law for a wavelengths of a few km)  for
stellar  masses.

The limits on neutron star heating by accretion of mirror baryons and other
intersing effects  
are discussed more quantitatively in the paper \cite{Sandin2008}.

\section{Microlensing}

  Invisible stars can be found by effect of microlensing. Those objects are called  now  {\sc macho}s.
Below I will write  {\sc macho} for ``Massive Astrophysical
Compact Halo Object'' and MACHO for the collaboration \cite{MACHO2000} studying them .
For mirror stars the effect of microlensing as a means to discover them was discussed in
\cite{Berezhiani1996,Blinnikov1998,BerCiarc2006}.
The difference with macrolensing is that the image is not resolved, the effect
is in the enhancement of the detected flux of the lensed object due to the passage
of the lens across the line of sight.

The MACHO results \cite{MACHO2000}
of photometry on 11.9 million stars in the Large Magellanic Cloud (LMC)
has revealed 13 - 17 microlensing events.
This is significantly more than the $\sim$ 2 to 4 events expected from lensing by known stellar populations.
If true, this points to the existence of stellar-like invisible objects in halo of Milky Way, perhaps
MM stars.

Not all DM in the halo may be in  {\sc macho}, only a fraction, usually denoted by
$f$.
MACHO group  (LMC) has given for the halo fraction $0.08<f<0.50$ (95\% CL)
to {\sc machos}  in the mass range $0.15M_\odot<M<0.9M_\odot$.
EROS collaboration  has placed only an upper
limit on the halo fraction, $f<0.2$ (95\% CL) for objects in this mass range.
Later they have given $f<0.1$ for $10^{-6}M_\odot<M<1M_\odot$ (see refs. in \cite{EROS2007}).
AGAPE collaboration (working on microlensing of M31 - Andromeda galaxy) finds
the halo fraction in the range $0.2<f<0.9$.
(See \cite{AGAPE2008} where arguments are given in favor of true halo {\sc macho}s against
self-lensing by ordinary stars in M31.)
The MEGA collaboration marginally conflicts with them finding  a halo fraction $f<0.3$
\cite{MEGA2007} .

Thus, the results of different groups are partly conflicting and quite confusing for a theorist.
Some workers even declared the
End of MACHO Era (1974-2004), e.g.  \cite{YooCG2004}.

Based on the results of MACHO group Bennett \cite{Bennett2005} has concluded that
{\sc macho}s are real and lensing optical depth is
$\tau=(1.0 \pm 0.3) \times 10^{-7}$ .
This is not far away from the old MACHO value of $\tau=1.2 \times 10^{-7}$,
 but the error is so large
that it is within $\la 2 \sigma$ of the effect from ordinary stellar populations alone  \cite{EvansBel2007}.
  Evans and  Belokurov  \cite{EvansBel2007}
confirm lower number (optical depth $\tau$) of {\sc macho}s to
the LMC from EROS \cite{EROS2007}
collaboration, who have reported $\tau  < 0.36 \times 10^{-7}$.
But there is a new paper of the same group -- in favor of {\sc macho}s from binary star
studies \cite{Quinn2009} putting arguments against the pessimistic conclusions of  \cite{YooCG2004}.

Most recent study with  HST (Hubble Space Telescope) follow-up shows
that MACHO collaboration data are not contaminated by
background events \cite{MACHO2009}, and this adds some optimism to those
who believe that invisible stars do exist.


Microlensing results do not allow us to have too many invisible stars.
Let us take their local density like  $8 \cdot 10^{-4} \msun/\mbox{pc}^3$
(solar masses per cubic parsec), i.e. {\it two orders of magnitude lower} than the number
$ 0.076 \pm 0.015 \msun/\mbox{pc}^3$ given by Hipparchos satellite from accurate studies of
dynamics of visible stars.

If about half of them are white dwarfs (this is quite likely in MM-sector) then we may have
{$\sim 2 \cdot 10^3$ mirror white dwarfs within 100 pc} -- just one order less than visible white
 dwarfs (Holberg et al. \cite{Holberg2008}) -- and the chances to find a few of them very close to us
appear nonzero.

\section{Limits on self-interacting DM}
\label{Limits}

Recently the cluster of galaxies 1E 0657-558 (``Bullet'' cluster) has drawn general attention.
It is an example of collision of two clusters, which gives
new clues to the proof of reality of DM and pauses
difficulties for  alternatives to DM like modified gravity  (MOND).
Here it is clear that the DM distribution, which is traced by the effect of lensing,
follows the distribution of stars and galaxies (which are effectively a collisionless
gravitating gas) and is shifted from hot X-ray emitting gas which dominates the baryon
mass in the cluster \cite{Clowe2004,Clowe2006,Bradac2006}.

Many physicists tend to use this example as a direct proof of a small cross-section
of self-interaction of DM particles.
In reality it proves only what is observed and nothing more:
DM behaves like ordinary stars which interact
only by gravity and form a collisionless matter.
But of course O-stars are made of strongly and electromagnetically interacting particles.

Let us take another example, the cluster  MACS J0025.4-1222 which is similar to the Bullet cluster.
It is found there that the DM  cross-section obeys the limit
$ \sigma/m<4 \; \mbox{cm}^2 \mbox{g}^{-1}$ \cite{Bradac2008}.
But this limit would be true for DM particles only if they do not form bound objects!
If DM there is anything like our normal solid bodies, say ice with density like
$1\; \mbox{g cm}^{-3} $  with the size $r$ larger than a few cm, then this limit is very well satisfied
($ \sigma $ is $\propto r^2$, while mass is $\propto r^3$).

So if the properties of DM particles are similar to our particles, and they are able to
form stars, planets, asteroids, etc. (like MM), then all observations of the merging clusters
are reproduced.
However, we have to squeeze a major fraction of MM particles into those
compact objects, not leaving a large fraction in the form of gas like we have
in OM in clusters of galaxies where the O-gas dominates in baryon component.

Dark matter in the Abell 520 cluster presents  another case,  difficult for CDM with
WIMPs.
In contrast to the Bullet cluster, the lensing
signal and the X-ray emission coincide here, and are away from  galaxies indicating
that DM is collisional like in O-gas:
``\ldots a mass peak without galaxies cannot be easily explained within the current collisionless dark matter paradigm'' \cite{Mahdavi2007}.
Thus Abell 520 cluster indicates a significant self-interaction cross-section.
In MM model we may get this leaving a major fraction of DM in form of gas.
``It is hard for the WIMP based
dark matter models to reconcile such a diverse behaviour. Mirror dark matter
models, on the contrary, are more flexible and for them diverse behaviour
of the dark matter is a natural expectation''  \cite{Silagadze2008}.

L.B.Okun said: ``MIRSY Dark Matter is richer than SUSY Dark Matter''.

Other interesting examples, like Hoag's object, are also discussed by Silagadze  \cite{Silagadze2008}.

\section{Conclusions}
\label{sec:Conc}

I can draw a few conclusions from the above sketchy notes, anyway, some of them
are important in general studies of Dark Matter candidates.

It is clear that the concepts of virial velocity or virial temperature are not always
relevant.

One should not use the limits on self-interaction cross-sections derived from observations
of colliding clusters as cross-sections for individual particles: they may form macroscopic bodies.

The effect of the heating of OM interstellar medium by MM neutron stars and white dwarfs
is very weak but may be observable.
One has to  search for weak but high proper motion peculiar objects, especially in radio-range.
Relation of these problems to
RadioAstron mission and LOFAR \cite{LOFAR2007} may be discussed and requires a detailed
modeling.
Peculiar binaries with low luminosity components, especially in star clusters, are also
interesting.

The effect of the heating of OM neutron stars by accretion of M-baryons is negligible.

There are several arguments in favor for MM (or other hidden sector DM) search:

\begin{itemize}
\item[--] {  Non-detection of WIMPs.}
\item[--] {  Unknown properties of DM particles with respect to clustering
 on stellar-size and mass scale.}
\item[--] {  Microlensing: lack of normal stars to explain {\sc macho} events.}
\item[--] {  Clusters of galaxies like Abell 520 which are hard to explain in pure CDM
 picture.}
\item[--] {  New kinds of mysterious transients like reported by
K.~Barbary {\em et~al.}, \cite{Barbary2009} may be explained by MM objects \cite{Sandin2008},
anyway it must be a very unusual accretion  event \cite{Soker2008}.}
\item[--] { One may speculate on possible MM life and intelligence  (L.B.Okun, N.S.Kardashev).}
\end{itemize}

 MM-baryons may form stellar-mass/size  compact objects, so they are interesting for
microlensing, however, the current situation with observational data on  {\sc macho}s
is not satisfactory: there are conflicting results of different groups.
It seems that total DM halo  mass in Galaxy cannot
be explained by invisible stars,  then more probable model for DM is a combination
CDM+MM, or WDM+MM, where
 WDM is Warm Dark Matter. WDM may be, e.g., a sterile neutrino
(with mass few keV) or gravitino \cite{Kusenko2007,Gorb2008a,Gorb2008b,Boya2008}.

Another window into the Mirror World may be
photon - mirror photon mixing.
There are  strong limits on this mixing:
orthopositronium experiments put an upper limit $ \epsilon < 1.55 \cdot 10^{-7}$ (90\% C.L.)
 \cite{Badertscher2007}.
Nevertheless, even smaller value of  $ \epsilon \sim 10^{-9}$
can explain DAMA/LIBRA experiments  \cite{Foot2008}.
 One should  analyze experiments like DAMA, taking into account this option,
 and if $\epsilon \ne 0$ look for
transformation of invisible mirror photons into visible ones.
 Other exotic  effects are possible for $\epsilon \ne 0$, e.g. MM $e^-e^+$ transformations
into OM $e^-e^+$  in M-pulsar,
or a mirror magnetar, looking   as a peculiar object with strong $B \sim 10^6$ G (D.Kompaneets).

I have not discussed  cosmological evolution of MM.
It is not clear a priori that pure MM can give sufficient amplitude of
perturbations to provide the ``DM boost'' for visible matter structure formation:
on the first glance, it seems that MM has a relatively short period of life
in radiation-dominated era (contrary to WIMPs) when there is a logarithmic growth of
perturbations.
Nevertheless, MM is able to replace Cold Dark Matter (CDM)
in evolution of perturbations \cite{Berezhiani1996,Ciar2005a,Ciar2005b}.
The question was analyzed in detail in  \cite{Ciar2005a,Ciar2005b} for various mixtures of
MM and CDM.
It is demonstrated  \cite{Ciar2005b} that the LSS (Large-Scale Structure) spectrum
is sensitive to MM parameters, due to oscillations in mirror baryons and
the collisional mirror Silk damping.
If  $T_{\rm MM} < 0.3 T_{\rm OM}$  in Early Universe then
M-baryons will decouple from M-radiation
before horizon crossing, and then as soon as they enter horizon, they
do not oscillate, but grow exactly as CDM would do in radiation-dominated universe.
Then CMB and LSS power spectra in  linear regime are equivalent for mirror and CDM cases.
For $T_{\rm MM} > 0.3 T_{\rm OM}$ the LSS spectra strongly
depend on the amount of mirror baryons, but for lower $T_{\rm MM}$  the entire dark matter
could be made of mirror baryons given current  observational limits on the CMB and LSS spectra.

Due to lower $T_{\rm MM}$ a larger fraction of mirror He is produced
in cosmological nucleosynthesis, than in OM, hence faster stellar evolution follows 
\cite{BerCiarc2006}.
It may result in lower fraction of MM intergalactic gas which is needed for explaining cases
similar to Bullet cluster.

\begin{acknowledgments}
This paper is written for a special issue of Yadernaya Fizika (Physics of Atomic Nuclei) dedicated to 80th birthday of L.B. Okun. I am very deeply grateful to Lev Borisovich for his lectures, books, joint work, and numerous discussions which always are a source of inspiration.

The paper is partly based on my talk at a conference
``Radio Universe at Ultimate Angular Resolution''  \cite{Blinnikov2008}
and I am grateful to N.S.Kardashev for his invitation and encouragement.
I thank M.I.Vysotsky,  V.Lukash,
D.Kompaneets, Z.Berezhiani, Z.Silagadze, A.Kusenko, P.Ciarcelluti, F.Sandin, for discussions and correspondence.
My work is supported by IPMU, University of Tokyo,
via World Premier International Research
Center Initiative (WPI), MEXT, Japan.
I am grateful to K.Nomoto  and H.Murayama for their kind hospitality.
The work in Russia is partly supported  by the Russian
Foundation for Basic Research grant RFBR 07-02-00830-a and by Russian
Scientific School Foundation under grants NSh-2977.2008.2.
NSh-3884.2008.2.
\end{acknowledgments}

\newpage

\centerline{REFERENCES}


\end{document}